# Blockchain, Fog and IoT Integrated Framework: Review, Architecture and Evaluation

Tanweer Alam[1], Mohamed Benaida[2]
Faculty of Computer and Information Systems, Islamic University of Madinah, Saudi Arabia[1,2.]

**Abstract**—In the next-generation computing, the role of cloud, internet, and smart devices will be capacious. Nowadays we all are familiar with the word smart. This word is used a number of times in our daily life. The Internet of Things (IoT) will produce remarkable different kinds of information from different resources. It can store and process big data in the cloud. The fogging acts as an interface between cloud and IoT. The IoT nodes are also known as fog nodes, these nodes are able to access anywhere within the range of the network. The blockchain is a novel approach to record the transactions in a sequence securely. Developing new blockchains based integrated framework in the architecture of the IoT is one of the emerging approaches to solving the issue of communication security among the IoT public nodes. This research explores a novel approach to integrate blockchain technology with the fog and IoT networks and provides communication security to the internet of smart devices. The framework is tested and implemented in the IoT network. The results are found positive.

## 1. Introduction

The proposed research is a step forward in wireless networking and IoT where we propose a new middleware framework based on fog and blockchain technology in IoT networks. The wireless communication is the key point for communicating two or more gadgets in IoT. The proposed research work in this study is a novel approach to create a new integrated model with fog, IoT and blockchain technologies. The research outcome is to establish a new integrated framework to solve the issue of communication security. The proposed research uses the correct and efficient simulation of the desired study and can be implemented in a framework of the Internet of Things. In the future, researchers can enhance this research and implement it on the internet of everything framework. Developing a new blockchains based middleware framework in the architecture of the Internet of Things can be a valuable framework to renovate the performance of IoT framework in a heterogeneous atmosphere. The wireless communication is the fastest growing research area that facilitates users to interact with each other without using wires. The Internet of Things is based on wireless networking at all. At the beginning of the Internet, it was developed to communicate one device to another device using accessing the browsers. However, in the current era, high speed smart efficient devices with many advanced technologies like low power consumption, etc. available to communicate with each other. The extension of fog in this framework works on physical things under IoT. The fogging explores the cloud to be closer to the physical objects [1]. The IoT nodes are also known as fog nodes, these nodes can have access anywhere within the range of the network. Fog computing reduces delay and saves bandwidth [2]. This study will use in the IoT framework. However, several frameworks are studied in the literature review. This research adds the enhanced blockchain and fog to develop efficient IoT framework for communication among smart devices. The comparison of this research with past investigations retransmission limits, affirmation, bundle length variety and worsen dissemination of packets generation is represented.

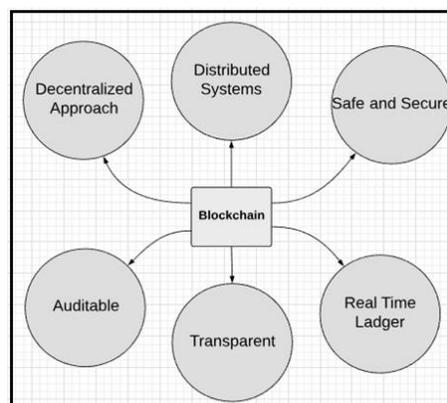

Figure 1. Features of Blockchain

The transactions in the proposed system are secured due to its private and public key generation. The IoT node operator can do the transaction with a new block in the system using authentication keys and transmits it to the neighborhood devices [3], [22], [23]. The authenticated Blockchain is not the point-to-point connection among the IoT nodes [4]. A device in the blockchain network is called miner who is responsible to mine a new block after solving the proof of work. This miner device is used to validate the transactions and valid transaction transmits into the blockchain network. This newly generated block should match the hash with its previous block. The IoT-based devices are worked with the Fog nodes locally first before sending the information [5]. The blockchain keeps a record of all transactions made by each device [6] (Figure 1).

The algorithm has been experimentally implemented. The proposed framework predicts well our comprehensive experiments. Further, validate the mathematical results. The objective of this research is to create a new model for communication among the internet of things and Fog computing. This research is based on blockchain technology with Middleware, Fog, and IoT. The framework can provide QoS through reducing the traffic rate vacillation also the variety of the number of smart devices. In this research, we consider the idle state in order to make our examination more efficient, at that point, the general execution regarding the overall performance of the framework is evaluated. The IoT-Fog in this framework will monitor and analyze the real-time data collected from fog nodes and then taking the action.

## 2. Backgrounds

The objective of this research is to create a communication framework and provides reliable, secure and fast connection using fog and blockchain among the smart devices in the internet of things (Figure 2). The previous studies have been focused on the creation and optimization of the framework for communication, but such research doesn't perform the full framework for IoT-Fog communication among the internet of smart devices. The proposed research plan builds research on extending the communication in the internet of things using fog and blockchain technology. The transfer data from one configuration to another using wireless networks start from 1973 in the form of the packet's radio network. They were able to communicate with another same configuration device. There are a lot of papers published on blockchain and internet of things in 2017-18 by the several authors but some of them such as [7], [8], [9], [10], [11], [12], [13], [14], [15], [16], [17] and [18] are related to our research. In 2015, Florian Tschorsch and Bjorn Scheuermann have presented the idea of bitcoin protocols, blockchain and its structure [7]. They also investigate the blockchain transaction security in networking. In 2016, Ali Dorri, Salil S. Kanhere, and Raja Jurdak have published an article about the blockchain technology in IoT [8]. In this paper, they have presented the architecture of blockchain technology for the internet of things to fire the transaction using a decentralized approach.

The fog and cloud are used as a host platform in the article [9] to deploy the blockchain in the internet of things. In 2016, Boohyung Lee and Jong-Hyuk Lee [10] were designed a blockchain architecture for the security of transactions in the IoT networks, these transactions were fired by the IoT-based smart devices [24].  In 2017, Alexandru Stanciu has published an article [11] on blockchain technology that is integrated with the IoT-nodes to execute the edge resources in the internet of things. The integration architecture of the software-defined network (SDN), fog, the blockchain, and IoT is designed in the article [12] to support security, availability, delivering the information, reduce the end-to-end delay among smart devices. The Blockchain-based transactions in IoT are discussed in the article [13]. In the article [14], a detailed description of the security of transactions in the blockchain and operating it in the internet of things are discussed and provide an approach to integrate both technologies together. In 2018, Runchao Han et.al. were published an article based on the evaluation of blocks in blockchains network of the internet of things. They were evaluated and tested the byzantine-tolerant blockchains adopted by the internet of things [15].  The integration of blockchain technology with the cloud and IoT is presented in [16]. Authors are focused on the sensor data security in this article [16]. In June 2018, Alfonso Panarello et. al. are published an article on the integration of blockchain and IoT [17]. They discussed the challenges and opportunities to integrate the blockchains with IoT and cloud. In October 2018, Babatunji Omoniwa et. al. are presented fog/edge-based IoT framework that enables the quick reply to the IoT nodes to request in Fog-IoT framework also, storage and computation services [18].

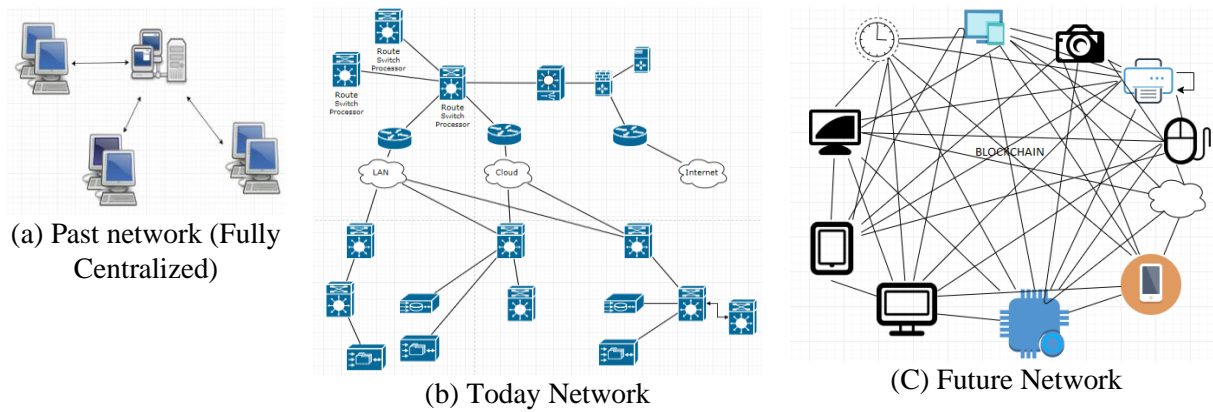

(a) Past network (Fully Centralized)

(b) Today Network

(C) Future Network

Figure 2.    Present, Past, and Future Networks

## 3. Methodologies and Architecture

The main contribution of this research is designing a framework for communication in the Internet of things using fog and blockchain technology. The proposed framework is specifically appropriate for applications in which data is periodically transmitted on the internet of the smart devices environment. In these applications, on one hand, packets are being produced based on a certain period time pattern. On the other hand, the service time is always a random variable with the general distribution. Therefore, service time might temporarily exceed the period time which, as an inevitable consequence some packets might encounter a busy channel and be dropped. We solve this problem by proposing the new middleware framework. We demonstrate that the proposed IoT-Fog framework [25], [26], [27], not only increases the throughput, but also the direct connection between the generation (sensors) and communication packet systems are eliminated which make the system far more stable. Moreover, in order to enhance the proposed model, we have employed a retransmission scheme, variable packet length, and saturated traffic condition. The solution to this research is summarized as follows. The implementation of IoT-Fog framework to communicate securely between the IoT nodes in 5G will be programmed to execute on to the IoT [28], [29], [30]. This research will implement into the three-layer model, these layers are Fog, Blockchain, and IoT. The proposed study supports wireless networking technology to establish an IoT-Fog framework among internet of things devices.

The proposed framework introduced and presented its role in IoT. The IoT-Fog framework has the following components:

a)      Devices (Things)
b)      Internet
c)      Middleware
d)      Fog computing with blockchain

In the fog, the modality servers stored secure resources, the proxies are third-party servers that can store protected data and the owners are the legitimate machines [31], [32]. The keys server in middleware generates encryption and decryption keys. The token provided to the smart device by the authorized blockchains database has the authority to accesses the framework, request the keys from a key server, fetch the data from the cloud. Figure 3 presents the IoT-Fog-blockchain framework.

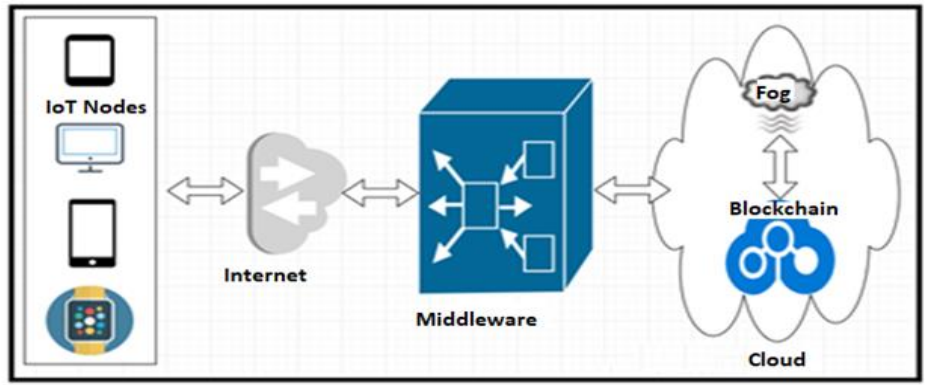

Figure 3. IoT-Fog-Blockchain Framework

The following steps are used in the proposed framework.

1. The smart contracts published by the modality servers, proxies and owners of fog in the authorized blockchains database.

2. The smart device of IoT finds the smart contacts from the authorized blockchains database.

3. The authorized blockchains database generates the token for the smart device of IoT [33].

4. The smart device requests the keys from a key server in middleware and sends the token with the request.

5. The key server in middleware verify the token from an authorized blockchain database and generates a key for the smart device and response back to the smart device [34].

6. Now the smart device of IoT is authorized to access the data from the cloud.

The framework itself has three major components: the IoT nodes, Fog, and blockchain.

The IoT node (IoTN) can be expressed by $\aleph$ with six tuples as follows [19].

$$\aleph = \{I_d, S, T, L, S_p, O\} \qquad .....................................(1)$$

Where $I_d$ is the unique identification of the IoT node, S is the status of IoT node, T is the type, L is the location, $S_p$ is the specification and O is the application instance.

Every IoT node has a unique identification number ($I_d$) provided by the IoT office after registering the device to the network. The $I_d$ is generated based on the inherent pattern of the node. The $I_d$ is the prerequisite in IoT nodes network. The Status of the IoT nodes could be inactive or an inactive state. The value of S can be 0 or 1. 0 means node is in an inactive form and 1 means node is inactive form. An IoT node can sense different types of events. Suppose $T_1$, $T_2$, $T_3$, ......, $T_n$ are the type of events that are sensed by the IoT node. Then $T = \{T_1, T_2, T_3, ....., T_n\}$. The location of the IoT node can be obtained as follows in the x-axis, y-axis, and z-axis during the time t [19].

$$L = \{L_X, L_y, L_z, t\} \qquad .........................(2)$$

Where $L_X, L_y, L_z$ are the location components of IoT node in x,y, z-axis at the time stamp t.

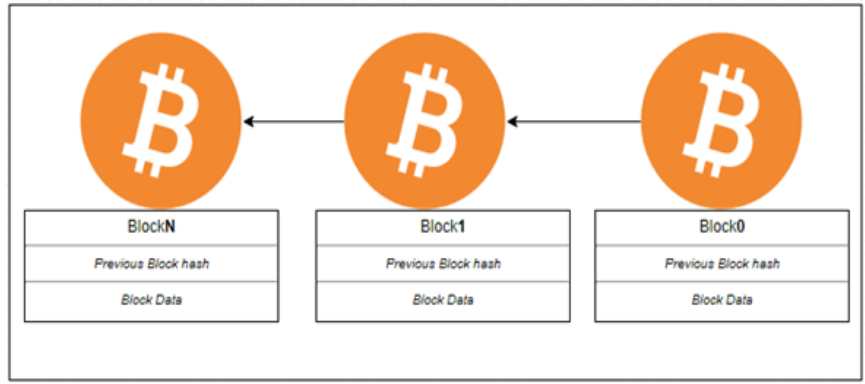

Figure 4. Blocks in Blockchain

Packets are transmitted in fog using blockchains node (BCN). Each BCN is also called block (Figure 4). Every block has its hash code (unique id like a fingerprint), previous block hash code and their own data. The BCN has a connection to exchange information (CEI). Every CEI contains a lot of blocks with its previous blocks hash code and data. These blocks are connected to each other with security through cryptography techniques. The BCNs are similar to the linked list node in the data structure. So, we can say that blockchain is a complex data structure. The blocks are distributing in a decentralized system using the point-to-point topological network. When a new block is created then it moves to the network and visit every connected BCN and checks its authentication. If it is valid then it will connect to the blockchain and its hash will generate only once. This newly generated block stores the hash of the previous block and connects to the chain [20].

The data transmission among the IoT-devices in the framework can passage from IoT-gateways (IoTG), fog gateways (FGW), blockchain gateway (BCG) or other gateways (OGW). The OGW is the combination of specified gateways other than the IoTG. The IoTG is the collection of gateways among the IoT devices [21]. Suppose that the communication delay among IoT devices is pondered immaterial. Consider $\Omega 1$, $\Omega 2$, $\Omega 3$ and $\Omega 4$ are the transmissions delay function among the IoT devices to OGW, OGW to IoTG, IoTG to FGW and FGW to BCG respectively. Suppose £1, £2, £3 and £4 are the latencies of OGW, IoTG, FGW, and BCG. Thus, the mean transmission latency can be obtained by the following function [21].

$$\sigma = (\Omega_1\mu + \Omega_2\theta + \Omega_3\tau + \Omega_4\epsilon) + (\pounds_1\mu + \pounds_2\theta + \pounds_3\tau + \pounds_4\epsilon) \ \dots\dots\dots\dots\dots\dots\dots\dots\dots\dots\dots\dots(3)$$

Where $\mu, \theta, \tau \ and \ \epsilon \ (\mu > \theta > \tau > \epsilon)$ are the total data packets sent by IoT devices, OGW, IoTG, and FGW. When data are transmitted from IoT device to the OGW and OGW to the IoTG then it requires energy consumption say $\vartheta_1$ and $\vartheta_2$ respectively. We consider $\vartheta_3$ and $\vartheta_4$ are the energy consumption from IoTG to the FGW and FGW to the BCG. Suppose $\omega_1$, $\omega_2$, $\omega_3$ and $\omega_4$ are the energies demands to evaluate the unit byte data received from OGW, IoTG, FGW, and BCG respectively. The total value of energy diffusion $(\varphi_t)$ in transmission can be obtained by the following formula.

$$\varphi_t = \left[\left\{\vartheta_1\sum\nolimits_{i=1}^{x}\sum\nolimits_{j=1}^{y}\gamma_{i,j} + \vartheta_2\sum\nolimits_{i=1}^{y}\sum\nolimits_{j=1}^{z}\alpha_{i,j} + \vartheta_3\sum\nolimits_{i=1}^{z}\sum\nolimits_{j=1}^{t}\beta_{i,j} + \vartheta_4\sum\nolimits_{i=1}^{t}\sum\nolimits_{j=1}^{w}\varepsilon_{i,j}\right\} + \left\{\omega_1\sum\nolimits_{i=1}^{x}\sum\nolimits_{j=1}^{y}\gamma_{i,j} + \omega_2\sum\nolimits_{i=1}^{y}\sum\nolimits_{j=1}^{z}\alpha_{i,j} + \omega_3\sum\nolimits_{i=1}^{z}\sum\nolimits_{j=1}^{t}\beta_{i,j} + \omega_4\sum\nolimits_{i=1}^{t}\sum\nolimits_{j=1}^{w}\varepsilon_{i,j}\right\}\right]$$

$$\dots\dots\dots\dots\dots\dots\dots\dots\dots(4)$$

Where $\sum\nolimits_{j=1}^{y}\gamma_{i,j}$, $\sum\nolimits_{j=1}^{z}\alpha_{i,j}$, $\sum\nolimits_{j=1}^{t}\beta_{i,j}$ and $\sum\nolimits_{j=1}^{w}\varepsilon_{i,j}$ are the total number of bytes transmitted from IoT device to OGW, from OGW to IoTG, from IoTG to FGW and from FGW to BCG at timestamp t.
The total rate of energy consumption ($\delta_{Fog}(t)$) at time-stamp t can calculate as follows.

$$\delta_{Fog}(t) = \frac{\varphi_t}{t}$$

So,

$$\delta_{Fog}(t) = \left. \left[ \begin{array}{l} \left\{ \vartheta_1 \sum_{i=1}^{x} \sum_{j=1}^{y} \gamma_{i,j} + \vartheta_2 \sum_{i=1}^{y} \sum_{j=1}^{z} \alpha_{i,j} + \vartheta_3 \sum_{i=1}^{z} \sum_{j=1}^{t} \beta_{i,j} + \vartheta_4 \sum_{i=1}^{t} \sum_{j=1}^{w} \varepsilon_{i,j} \right\} + \\ \left\{ \omega_1 \sum_{i=1}^{x} \sum_{j=1}^{y} \gamma_{i,j} + \omega_2 \sum_{i=1}^{y} \sum_{j=1}^{z} \alpha_{i,j} + \omega_3 \sum_{i=1}^{z} \sum_{j=1}^{t} \beta_{i,j} + \omega_4 \sum_{i=1}^{t} \sum_{j=1}^{w} \varepsilon_{i,j} \right\} \end{array} \right] \middle/ t \right.$$

…………………………………… (5)

The performance of the proposed system is evaluated through different experiments. Firstly, I have created thousands of blocks with a fixed size by using open source software (Node.js). I have created the IoT network, connected with cloud and also, created fog and blockchain.

## 4. Results and Discussion

Our proposed system is now ready to evaluate the performance. I found the experimental results positive compared to previous studies. In the blockchain, I need miners for a different experiment. Firstly, I have selected two miners with fixed transactions (suppose 5) and set the fog demands for each miner (ex. 30,50). Secondly, I have selected three miners with fixed transactions (suppose 5) and set the fog demands for each miner (ex. 30,40,50).

Table 1: Evaluation Data

| Miners node | Transmission delay (MS) | Action duration (S) | Energy consumption (KJ) |
|---|---|---|---|
| 5 | 146 | 301 | 5.3 |
| 5 | 12 | 906 | 5.3 |
| 5 | 100 | 1220 | 5.7 |
| 10 | 15 | 320 | 12.5 |
| 10 | 12 | 505 | 15.8 |
| 10 | 54 | 230 | 11.8 |
| 50 | 63 | 245 | 20.9 |
| 50 | 43 | 600 | 25.5 |
| 50 | 6 | 309 | 32.2 |
| 100 | 400 | 405 | 35 |
| 100 | 733 | 506 | 34.5 |
| 100 | 245 | 915 | 36.8 |

If the fog demands are increasing then the possibility of mining the block by the miner is higher. The computational requirements such as processor usage, Memory usage in blockchain compare to fog and cloud are evaluated (Figure 5, 6). I found that the processor usage in blockchain was lower than the fog and cloud.

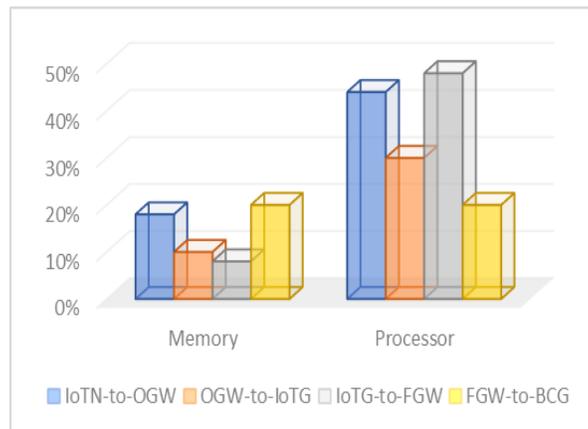

Figure 5. Computational requirements

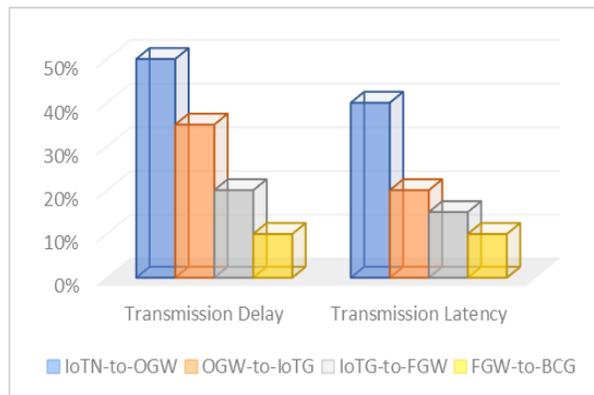

Figure 6. Transmission delay and latency computation

The memory usage depends on the number of blocks and the number of transactions. If a number of blocks in the chain increases then memory usage also increases (Figure 7). Similarly, if the number of transactions increases then the memory usage also increases.

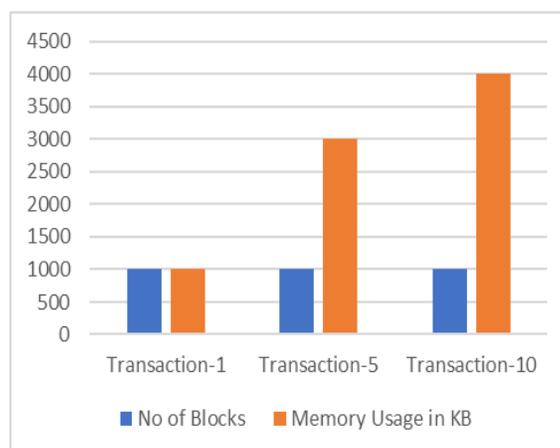

Figure 7. Memory usage and number of blocks vs. transactions

I have calculated the transmission delay for synchronization a block in blockchain in this study (Figure 8).

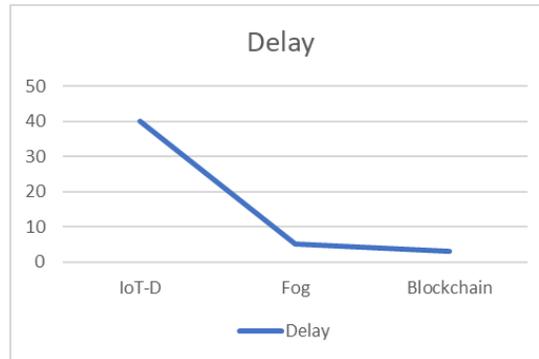

Figure 8. Transmission delay for synchronization a block

I have tested the proposed framework using many tests. The summary is displayed in the following table.

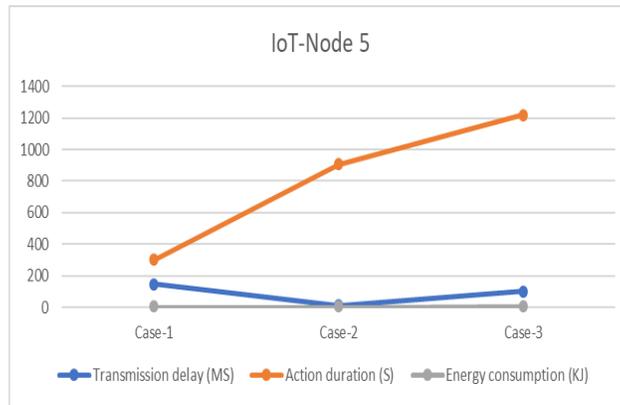

Figure 9(a): Performance on 5 IoT nodes

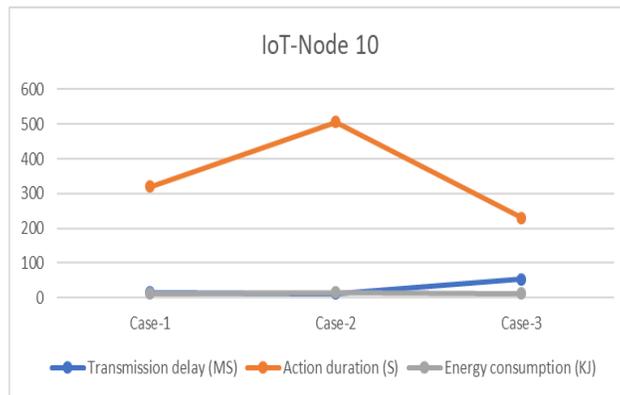

Figure 9(b): Performance on 10 IoT nodes

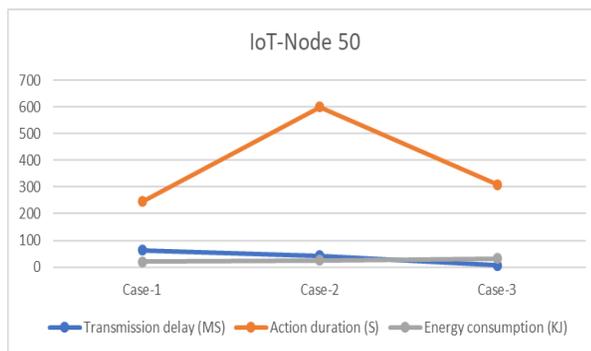

Figure 9(c): Performance on 50 IoT nodes

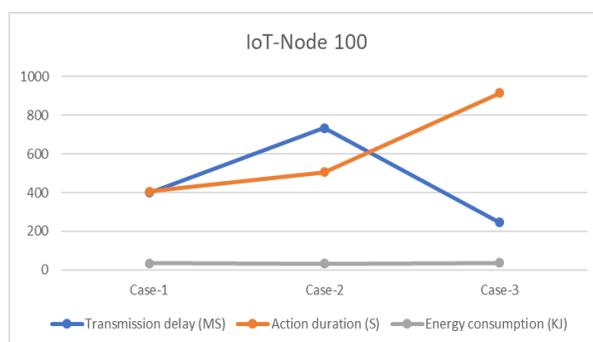

Figure 9(d): Performance on 100 IoT nodes

In figure 9(a), 9(b), 9(c) and 9(d) represent the performance of IoT-Nodes in the proposed framework.

## 7. Conclusion

The proposed framework is designed and implemented using IoT, fog and blockchain technologies. This study can be a valuable framework to improve the performance of IoT framework in a heterogeneous environment. This framework is appropriate for providing communication security where huge data is transmitted in a heterogeneous environment in the future. I have tested the system in different scenarios such as memory and processor usage in the integrated system and its impact on the performance of overall the system. I found that the proposed framework, not only increases the throughput but also the direct connection among IoT nodes are eliminated which make the system far more stable. The outcomes of this research established a new IoT framework with blockchain technology. In the future, researchers can enhance this research and implement it in the internet of everything with blockchain framework.